\documentclass[conference]{IEEEtran}
\usepackage[american]{babel}
\usepackage{graphicx}
\usepackage[utf8]{inputenc}
\usepackage{siunitx}
\usepackage{amsmath,amssymb,amsfonts,amstext}
\usepackage{ mathrsfs }
\usepackage{comment}
\usepackage{color}

\renewcommand{\abstractname}{Abstract}

\begin{document}
\def\IEEEkeywordsname{Keywords}

\title{SARS-CoV-2, a Threat to Privacy? \\ 
{\large (Originally submitted as a part of \#WirVsVirus Hackathon 21-22/03/2020)}
}
%
%
%

\makeatletter
\newcommand{\newlineauthors}{%
  \end{@IEEEauthorhalign}\hfill\mbox{}\par
  \mbox{}\hfill\begin{@IEEEauthorhalign}
}
\makeatother

\author{
\IEEEauthorblockN{Tim Daubenschütz}
\IEEEauthorblockA{
tim@daubenschuetz.de}
\and
\IEEEauthorblockN{Oksana Kulyk}
\IEEEauthorblockA{
oksana.kulyk@protonmail.com}
\and
\IEEEauthorblockN{Stephan Neumann}
\IEEEauthorblockA{
stephanneumann@tutamail.com}
\and
\IEEEauthorblockN{Isabella Hinterleitner}
\IEEEauthorblockA{
hinterleitner@techmeetslegal.at}
\and
\IEEEauthorblockN{Paula Ramos Delgado}
\IEEEauthorblockA{
ramosdelgado.paula@gmail.com}
\and
\IEEEauthorblockN{Carmen Hoffmann}
\IEEEauthorblockA{\textit{University of Innsbruck},\\
Innsbruck, Austria\\
carmen.hoffmann@protonmail.com}
\and
\IEEEauthorblockN{Florian Scheible}
\IEEEauthorblockA{\textit{Umit - Private University},\\
Hall in Tirol, Austria\\
mail@florian-scheible.com}
}

\maketitle

\begin{abstract}
\boldmath
 The global SARS-CoV-2 pandemic is currently putting a massive strain on the world's critical infrastructures. With healthcare systems and internet service providers already struggling to provide reliable service, some operators may, intentionally or unintentionally, lever out privacy-protecting measures to increase their system's efficiency in fighting the virus. Moreover, though it may seem all encouraging to see the effectiveness of authoritarian states in battling the crisis, we, the authors of this paper, would like to raise the community's awareness towards developing more effective means in battling the crisis without the need to limit fundamental human rights. To analyze the current situation, we are discussing and evaluating the steps corporations and governments are taking to condemn the virus by applying established privacy research.

\unboldmath
\end{abstract}

\begin{IEEEkeywords}
COVID-19, SARS-CoV-2, pandemic, privacy, security, critical infrastructure
\end{IEEEkeywords}

\section{Introduction}
Due to its fast spreading throughout the world, the outbreak of SARS-CoV-2 has become a global crisis putting stress on the current infrastructure in some areas in unprecedented ways, making shortcomings visible. Since there is no vaccination against SARS-CoV-2, the only way to deal with the current situation are nonpharmaceutical interventions (NPI's), to reducing the number of new infections and flatten the curve of total patients.

Having a look at European states like Italy, Spain, France, or Austria which, are in lockdown as of March 2020, keeping people away from seeing each other, their right to living a self-determined life is not in their hands anymore. As shown by Hatchett et al., this method showed a positive effect in St. Luis during 1918's influenza pandemic \cite{st_luis}, 
Nevertheless, its long-term effects on the economy and day-to-day life, including psychological effects on people forced to self-isolate, are often seen as a cause of concern \cite{psychological_impact}. 
Furthermore, some models show the possibility of a massive rise of new infections after the lockdown is ended \cite{NPI_reduce}. 

Hence, to handle the situation, measures are being discussed, some of which may invade a citizen's privacy. We can see an example of this approach in Asian countries e.g., South Korea \cite{south_korea} 
and Singapore \cite{singapore} 
where, besides extensive testing, methods such as tracing mobile phone location data in order to identify possible contact with infected persons  \cite{NPI_china}.

Other countries have taken similar measures. For instance, Netanyahu, Israel's Prime Minister, ordered Shin Bet, Israel's internal security service, 
to start surveilling citizens' cellphones \cite{israel_cellphone, israel_cellphone2, israel_cellphone3}. Persons who have been closer than two meters to an infected person are receiving text messages telling them to go into immediate home isolation for 14 days. As Shin Bet mandate is to observe and fight middle easter terrorism, naturally, Israel's citizens are now concerned that it is now helping in a medical situation \cite{israel_cellphone,israel_cellphone2,israel_cellphone3}. 
Within the EU, in particular, in Germany and Austria,  telecommunications providers are already providing health organizations and the government with anonymous data of mobile phone location data  \cite{A1_austria}.

Although nobody has evaluated the effectiveness of these measures, they raise concerns from privacy experts as the massive collection of data can easily lead to harming the population and violating their human rights if the collected data is misused.
In this paper, we discuss the privacy issues that can arise in times of crisis and take a closer look into the case of the German Robert Koch Institute receiving data from Telekom. We conclude by providing some recommendations about ways to minimize privacy harms while combating the pandemic.
\section{Privacy}
In this section we outline the general definitions of privacy, including describing the contextual integrity framework for reasoning about privacy, and discuss privacy harms that can occur from misuse of personal data. We furthermore discuss the issues with privacy that can occur during a crisis such as this global pandemic and what can be done to ensure information security and hence appropriate data protection.

\subsection{Definitions of Privacy}
Privacy is a broad concept which has been studied from the point of view of different disciplines, including social sciences and humanities, legal studies and computer science. The definitions of privacy are commonly centered around seeing privacy as \emph{confidentiality} (preventing disclosure of information), \emph{control} (providing people with means to control how their personal data is collected and used) and \emph{transparency} (ensuring that users are aware of how their data is collected and used, as well as ensuring that the data collection and processing occurs in a lawful manner) \cite{troncoso2020}.

Hannah Arendt, a Jewish philosopher who grew up in Germany in the beginning of the 20th century defined privacy within the context of public and private space. Her claim was that if there exists public space, there is also private space. Arendt considers the privacy concept as “distinction between things that should be shown and things that should be hidden”  \cite{arendt1958}. And that, private spaces exist in opposition to public spaces. Meaning, while the public square is dedicated to appearances, the private space is devoted to the opposite — to hiding and privacy. She associated privacy with the home. Due to the fact that we have become used to a "digital private space", such as our own email inbox or personal data on the phone, people are concerned and offended when the private, hidden space is violated. However, in times of crisis the term hidden or privacy becomes a new meaning.

Helen Nissenbaum, a Professor of Information Science, proposed the concept of \emph{contextual integrity} as a framework to reason about privacy. According to her framework, privacy is defined as \emph{adhering to the norms of information flow} \cite{nissenbaum2004}. These norms are highly contextual: for example, it is appropriate for doctors to have access to the medical data of their patients, but in most cases it is inappropriate for employers to have access to medical data of their workers. Nissenbaum distinguishes between the following five principles of information flow \cite{nissenbaum2019}: the \emph{sender}, the \emph{subject}, the \emph{receiver}, the \emph{information type} and the \emph{transmission principle} (e.g. whether confidentiality has to be preserved, whether the data exchange is reciprocal or whether consent is necessary and/or sufficient for the appropriateness of the data exchange). The norms governing these parameters are furthermore evaluated against a specific context, including whether the information flow is necessary for achieving the purpose of the context.

\subsection{Privacy Harms}
Data misuse can lead to different kinds of harms that jeopardise physical and psychological well-being of people as well as the overall society (see e.g. Solove, 2008). One of them is persecution by the government -- this might not be a big concern in democratic societies, but democratic societies can move into more authoritarian governance styles, especially is crisis situations. Even if this does not happen, there are other harms, e.g. a so called "chilling effect", where people are afraid to speak up against the accepted norms when they feel that they are being watched. Furthermore, harms can result from data leaks, like unintentional errors or cyberattacks. In these cases, information about individuals may become known to unintended targets. This can result in physical harm, stalking and damage of the data subject's personal relationships. Knowledge about one's medical data can lead to job discrimination. Leaked details about one's lifestyle can lead to raised insurance rates. Leakage of location data, in particular, can reveal a lot of sensitive information about an individual, such as the places they visit, which might in turn result in dramatic effects when revealed. Just think of closeted homosexuals visiting a gay clubs or marginalized religious minorities visiting their place of worship. Even beyond these concerns, access to large amounts of personal data can be used for more effective opinion and behavior manipulation, as evidenced by the Cambridge Analytica scandal \cite{cambridge_analytica}.

In summary, absence of privacy has a dramatic effect on our freedom of expression as individuals and on the well-functioning of the society as a whole. It is therefore important to ensure that the damage to privacy is minimized even in times of crisis.

\subsection{Privacy in Times of Crisis}
When we are considering the example of doctors treating their patients, we can use the framework of contextual integrity to reason about the appropriate information flow as follows: the patient is both the sender and the subject of the data exchange, the doctor is the receiver, the information type is the patient's medical information, the transmission principle includes, most importantly, doctor-patient confidentiality aside from public health issues. The overall context is health care, and the purpose of the context is both healing the patient and protecting health of the population. It can therefore be argued that in case of a global pandemic, one should allow the exchange of patient's data, especially when it comes to data about infected patients and their contacts, to the extent that it is necessary to manage the pandemic.

There is, however, a danger of misusing the collected data outside of the defined context -- the so-called "mission creep", which experts argue was the case with NSA collecting data from both US and foreign citizens on an unprecedented scale as an aftermath of the 9/11 terrorist attack \cite{schneier2013}. Furthermore, aside from the danger of collecting data by the government, the crisis situation leads to increase of data collection by private companies, as people all over the world switch to remote communication and remote collaboration tools from face-to-face communications. The data collection and processing practices of these tools,  however, are often obscure from their users: as known from research in related fields, privacy policies are often too long, obscure, and complicated to figure out, and shorter notices such as cookie disclaimers tend to be perceived as too vague and not providing useful information \cite{schaub2015design,kulyk2018website}. This leads to users often ignoring the privacy policies and disclaimers, hence, being unaware of important information about their data sharing. Moreover, even among the privacy-concerned users, the adoption of more privacy-friendly tools can be hindered by social pressure and network effects, if everyone else prefers to use more popular tools that are less inclined to protect the privacy of their users (as seen in studies on security and privacy adoption in other domains, see e.g. \cite{abu2017obstacles,volkamer2015socio}). This data collection even furthers the effects of the so-called \emph{surveillance capitalism}  \cite{zuboff2015}, which leads to corporations having even more power over people than before the crisis. This access to personal data by corporations is furthermore aggravated by an increased usage of social media platforms, increases in users sharing their location data and giving applications increased access to their phone's operating system. Lowered barriers and increased online activity that can be directly linked to an individual or an email address is a treasure trove for for-profit corporations that monetize consumer data. Many corporations are now getting free or low cost leads for months to come. 

A question that is often open for discussion is to which extent people themselves would be ready to share their data, even if it results in a privacy loss. As such, data sharing habits in general have been the topic of research, leading to discussions on so-called privacy paradox: people claiming that privacy is important to them, yet not behaving in a privacy-preserving way. The privacy paradox can be explained by different factors \cite{solove2020myth}. One of them is the lack of awareness about the extent of data collection as well as about the possible harms that can result from unrestricted data sharing. A further factor stems from decision biases, such as people's tendency to underestimate the risks that may happen in the future compared against immediate benefit. Another noteworthy factor are the manipulations by service providers (so-called dark patterns) nudging users into sharing more of their data contrary to their actual preferences. But rational decisions in times of crisis are even more difficult. Given the state of stress and anxiety many are in, people might be more likely to accept privacy-problematic practices if they are told that these practices are absolutely necessary for managing the crisis -- even if this is not actually the case.

The problem that people are more likely to surrender their privacy rights if they have already had to surrender other fundamental rights (such as freedom of movement due to lockdown restrictions) is reminiscent of the psychological mechanism of door-in-the-face technique. The door-in-the-face technique is a method of social influence, where we ask a person at first to do something that requires more than they would accept. Afterward, we ask for a second smaller favor. Research has shown that the person is now more likely to accept the other smaller favor \cite{moser2020}. In the case of the SARS-CoV-2 pandemic, governments first asked their citizens to self-isolate ( limiting significant fundamental freedom) before following up with the smaller favor of handing over some private data to fight the outbreak. However, according to Cantarero et al., the level of acceptance differs from individual to individual \cite{cantarero2017}, which makes it even more critical to rising consciousness in population.

At the same time, timely access to data voluntarily shared by people (in addition to the data collected by hospitals and authorities) can indeed help combat the epidemics. In this, we are supporting informed consent of data subjects, because it ensures that people will only share data with institutions that kept their data safe against privacy harms.

\subsection{Information Security Concerns}
In an increasingly digital world, establishing proper information security safeguards is critical in preventing data leaks, and hence, in preserving the privacy of data subjects. However, the situation of such a global pandemic places significant challenges on established workflows, information technology, and security as well, resulting in various issues. 

These problems arose when people stopped traveling, going into the office, and started working from home. While some companies and institutions have provided a possibility for remote work also before the crisis, or are at least infrastructurally and organizationally prepared, many are unprepared for such a dramatic increase of home office work. 
They face significant technical and organizational challenges, such as ensuring the security of their systems given the need for opening the network to remote access, e.g., via the so-called demilitarized zone (DMZ), or perimeter control, an extension of technical monitoring of the system and overall extension of system hardening is "hostile" (home) environments. A recent poll revealed that the security teams of 47\% of companies did not have "emergency plans in place to shift an on-premise workforce to one that is remote" \cite{homeoffice_cyberattacks}. 
Even worse, these challenges are more present in regulated (and therefore often critical) industries as Sumir Karayi, CEO and founder of 1E, in a Threatpost interview states:

\textit{\grqq Government, legal, insurance, banking and healthcare are all great examples of industries that are not prepared for this massive influx of remote workers [...] Many companies and organizations in these industries are working on legacy systems and are using software that is not patched. Not only does this mean remote work is a security concern, but it makes working a negative, unproductive experience for the employee. [...] Regulated industries pose a significant challenge because they use systems, devices or people not yet approved for remote work [...] Proprietary or specific software is usually also legacy software. It’s hard to patch and maintain, and rarely able to be accessed remotely.\grqq} \cite{work_from_home}

In consequence, the urgent need to enable remote collaboration related to the lack of preparation and preparation time may lead to hurried and immature remote work strategies.

At the same time, ensuring proper security behavior of the employees -- something that was a challenge in many companies also before the crisis -- is becoming an even more difficult task. We can currently see employees trying to circumvent corporate restrictions by sending or sharing data and documents over private accounts (shadow IT). Additionally, there is a surge of social engineering attacks among other phishing email campaigns, business email compromise, malware, and ransomware strains, as Sherrod DeGrippo, senior director of threat research and detection at Proofpoint, states \cite{cyber_security_threat}. 

Similar findings are provided by Atlas VPN research, which shows that several industries broadly use unpatched or no longer supported hardware or software systems, including the healthcare sector \cite{us_fight_covid}.

Together with immature remote strategies, information security and privacy risks may significantly increase and undermine the standardized risk management process. 

\subsection{General Data Protection Regulation (GDPR) in Context of the Pandemic}
The European Data Protection Board (EDPB) has formulated a statement on the processing of personal data in the context of the SARS-CoV-2 outbreak \cite{eu_statement}.

According to EDPB, data protection rules do not hinder measures taken in the fight against the coronavirus pandemic. Even so, the EDPB underlines that, even in these exceptional times, the data controller and processor must ensure the protection of the personal data of the data subjects.

Therefore, several considerations should be taken into account to guarantee the lawful processing of personal data, and in this context, one must respect the general principles of law. As such, the GDPR allows competent public health authorities like hospitals and laboratories as well as employers to process personal data in the context of an epidemic, by national law and within the conditions set therein.

Concerning the processing of telecommunication data, such as location data, the national laws implementing the ePrivacy  Directive must also be respected.
The national laws implementing the ePrivacy Directive provide that the location data can only be used by the operator when they are made anonymous, or with the consent of the individuals. If it is not possible to only process anonymous data, Art. 15 of the ePrivacy Directive enables the member states to introduce legislative measures pursuing national security and public security.

This emergency legislation is possible under the condition that it constitutes a necessary, appropriate, and proportionate measure within a democratic society. If a member state introduces such measures, it is obliged to put in place adequate safeguards, such as granting individuals the right to a judicial remedy. 

\section{Robert Koch Institute and Telekom Case}
In this section, we conduct a preliminary analysis of German disease control receiving movement data from a telecommunication provider.

In Germany, the authority for disease control and prevention, the Robert Koch Institute (RKI), made headlines on March 18, 2020, as it became public that telecommunication provider Telekom had shared an anonymized set of mobile phone movement data to monitor citizens' mobility in the fight against SARS-CoV-2. In total, Telekom sent 46 million customer's data to the RKI for further analysis. The German Federal Commissioner for Data Protection and Freedom of Information, Ulrich Kelber, overseeing the transfer, commented on the incident that he is not concerned about violating any data protection rules, as the data had been anonymized upfront \cite{sueddeutsche_telekom}.

However, researchers have shown that seemly anonymized data sets can indeed be "deanonymized" \cite{ohm2009}. Constanze Kurz, an activist, and expert on the subject matter, commented that she was skeptical about the anonymization. She urged Telekom to publicize the anonymization methods that were being used and asked the Robert Koch Institute to explain how it will protect this data for unauthorized third-party access. Several research studies had shown the deanonymization for data sets to extract personal information, including a case from 2016,  when a journalist and a data scientist acquired an anonymized dataset with the browsing habits of more than three million German citizens \cite{nature,isaca_journal}.

As at the moment, it is hard to tell whether disclosure of personal data is possible from the shared set (even more so given the development of new re-identification methods, including possible future development), we look at the worst-case scenario, namely, that personal data is deanonymize-able. Given this scenario, we use Nissenbaum's contextual integrity thesis to understand if privacy Telekom has violated its customer's privacy \cite{nissenbaum2019}. We do so by stating the context of the case, the norm -- what everyone expects should happen -- plus five contextual parameters to further analyze the situation. Table \ref{table:1} summarises the contextual integrity framework as applied to the German data sharing situation.
\begin{table}[h!]
\centering
\caption{Nissenbaum's contextual integrity applied to the Robert Koch Institute and Telekom case.}
\begin{tabular}{ | m{5em} | m{20em} | } 
 \hline
 Parameters & Contextual information \\
 \hline\hline
 Context & Health care, including public health \\ 
 \hline
 Norm & The Robert Koch Institute is responsible to prevent the spread of disease in Germany and is currently fighting further spread of SARS-CoV-2 \\
 \hline
 Data Subjects & 46 million customers of German Telekom \\
 \hline
 Sender & German Telekom or subsidiary Motionlogic \\
 \hline
 Recipient & The Robert Koch Institute \\ 
 \hline
  Information type & Mobile phone movement data \\ 
 \hline
   Transmission principle & Sender and Recipient are working with the German Federal Commissioner for Data Protection and Freedom of Information to ensure that the shared data was anonymized and is only used to prevent the spread of SARS-CoV-2 in Germany \\ 
 \hline
\end{tabular}
\label{table:1}
\end{table}

A principle that is perhaps most interesting for further elaboration is the transmission principle. Given the context and urgency of the situation, one might agree that having the German Federal Commissioner for Data Protection and Freedom of Information oversee the transaction and taking some measures to anonymize the data set appropriately serves as a practical solution towards limiting the spread of SARS-CoV-2, also without explicitly obtaining consent from data subjects. We do, however, assume that appropriate use of data would be limiting it to a specific purpose of combating the pandemic, and not reusing it to other purposes without further assessment. Note, however, that there is space for discussion, in which the community should be engaged, about the norms that apply in this situation, especially given the extraordinary situation and the severity of the crisis.

A further step of the contextual integrity is, however, also part of the contextual integrity framework to Nissenbaum's five parameter thesis of contextual information to create hypothetical scenarios that could threaten the decision's future integrity. We, therefore, consider the following hypotheticals, which we believe would violate contextual integrity:
\newline
\newline
\emph{Hypothetical scenario 1: "The Robert Koch Institute does not delete the data after SARS-CoV-2 crisis"}
\newline
\newline
\emph{Hypothetical scenario 2: "The Robert Koch Institute forwards data to other state organs or to third parties"}
\newline
\newline
\emph{Hypothetical scenario 3: "The Robert Koch Institute uses data for other purposes different from fighting SARS-CoV-2 spread or other similar public health crises"}
\newline
\newline

These hypothetical scenarios would violate the transmission principle that the data is only going to be used to handle the crisis (and, in the second hypothetical, also the receiver of the data). We believe a future assessment is necessary to determine if the data transfer was indeed necessary to fight the pandemic. Alternatively, if alternatively, customer permissions should have been required upfront. 
\newline
\newline
\emph{Hypothetical scenario 4: "The Robert Koch Institute requests data about phone calls and text messages exchanged by Telekoms' customers}
\newline
\newline
\emph{Hypothetical scenario 5: "The Robert Koch Institute requests data about Telekom customer movements from the last ten years}
\newline
\newline
These scenarios change the type of information. We want to argue that the new exchanged data no longer serves the purpose of fighting the pandemic. This point was also made by the Electronic Freedom Frontier organization \cite{eff}, noting that since the incubation period of the virus is estimated to last 14 days, getting access to data that is much older than that would be a privacy violation. We think that, similar to the first three scenarios, a further assessment, based on transparent information, is necessary.
\newline
\newline
\emph{Hypothetical 6: "The Robert Koch Institute uses the data purely for fighting SARS-CoV-2, but fails to keep it secure against hackers."}
\newline
\newline
As with the first three scenarios, the transmission principle of confidentiality is violated in this scenario, albeit unintentionally, and, in case of improper anonymization, personal information might still leak. Hence, a privacy violation has taken place. Referring to outlined information security concerns, an increase in cyber attacks related to improper information security management in a time of crisis significantly increases the risk. Given the above-outlined hypotheticals, we recommend implementing appropriate protection measures.T
\section{Recommendations and Outlook}
Countries around the world have already taken numerous initiatives to slow down the spread of the SARS-CoV-2, such as remote working, telemedicine, and online learning and shopping. That has required a legion of changes in our lives. However, as mentioned in previous sections, these activities come with associated security and privacy risks. 
Various organizations are raising concerns regarding these risks (see e.g., the statement and proposed principles from the Electronic Freedom Frontier \cite{eff}.

Of particular interest is the case of healthcare systems, which must be transparent with the information related to patients, but cautious with the disclosed information. Equally, hospitals might also decide to withhold information in order to try to minimize liability. That is a slippery slope: both cases -- no information or too much information -- might lead to a state of fear in the population and a false sense of security (i.e., no information means there is no problem) or a loss of privacy when we decide to disclose too much information.

In the current situation and others that might arise, principles, and best practices developed before the crisis are still applicable—namely, privacy by design principles, and most importantly, data minimization. Only strictly necessary data needed to manage the crisis should be collected and deleted once humanity has overcome the crisis.

In this context, patient data should be collected, stored, analyzed, and processed under strict data protection rules (such as the General Data Protection Regulation GDPR) by competent public health authorities, as mentioned in the previous chapter \cite{eu_statement}. An example of addressing the issues of data protection during the crisis can also be seen within the Austrian project VKT-GOEPL \cite{network_activity}. 
It was the project's goal to generate a dynamic situational map for ministries overseeing the crisis. Events, such as terrorist attacks, flooding, fire, and pandemic scenarios, were selected. Already ten years ago, the need for geographical movement data provided by telecommunication providers was treated as a use-case. Furthermore, the project initiators prohibited the linking of personal data from different databases in cases where this data was not anonymized. They recommended that ministries are transparently informing all individuals about the policies which apply to the processing of their data.

Regarding data analysis, we recommend that citizens only disclose their data to authorized parties, once these are putting adequate security measures and confidentiality policies in place. Moreover, only data that is strictly necessary should be shared. We think that proper data storage should make use of advanced technology such as cryptography. Patient data -- including personal information such as contact data, sexual preferences or religion amongst others -- should not be revealed. As anonymizing data has been shown to be a non-trivial task that is hard to achieve in a proper way, advanced solutions such as cryptographic techniques for secure multiparty computation or differential privacy algorithms for privacy-preserving data releases should be used.

Besides, to ensure privacy from the collection stage, consistent training of the medical personnel, volunteers, and administrative staff should be done. The current lack of training (due to limited resources, shortage of specialists, and general time pressure) leads to human errors and neglect of proper security and privacy protection measures.

A further concern, which we did not investigate in this paper is to ensure fairness when it comes to algorithmic decision making. As such, automated data systems ("big data" or "machine learning") are known to have issues with bias-based e.g., on race or gender that can lead to discrimination \cite{o2016weapons2}. In order to prevent such adverse effects during the crisis, these systems should furthermore be limited in order to limit bias based on nationality, sexual preferences, religion, or other factors that are not related to handling the pandemic.

Finally, we recognize that having access to timely and accurate data can play a critical role in combating the epidemic. Nevertheless, as discussed in previous sections, ignoring issues around the collection and handling of personal data might cause serious harm that will be hard to repair in the long run. Therefore, as big corporations and nation-states are collecting data from the world's population; it is of crucial importance that this data is handled responsibly and keeping the privacy of the data subjects in mind.

\bibliographystyle{IEEEtran}

\bibliography{artikel.bib}

\end{document}